\documentclass[prx,aps,amsfonts,amsmath,amssymb,twocolumn,superscriptaddress]{revtex4-2}
\usepackage{graphicx}
\usepackage{dcolumn}
\usepackage{appendix}
\usepackage{amsmath}
\usepackage{color}
\usepackage{bm}
\usepackage{multirow}
\usepackage{float}

\AtBeginDocument{%
	\newwrite\bibnotes
	\def\bibnotesext{aTe.bib}
	\immediate\openout\bibnotes=\jobname\bibnotesext
	\immediate\write\bibnotes{@CONTROL{REVTEX42Control}}
	\immediate\write\bibnotes{@CONTROL{%
			apsrev42Control,author="08",editor="1",pages="1",title="0",year="1"}}
	\if@filesw
	\immediate\write\@auxout{\string\citation{apsrev42Control}}%
	\fi
}%

\begin{document}

\title{Observation of a dynamic magneto-chiral instability in photoexcited tellurium}

\author{Yijing Huang}
\thanks{These authors contributed equally to this work.}
\affiliation{Department of Physics, The Grainger College of Engineering, University of Illinois Urbana-Champaign, Urbana, Illinois 61801, USA}
\affiliation{Materials Research Laboratory, The Grainger College of Engineering, University of Illinois Urbana-Champaign, Urbana, Illinois 61801, USA}

\author{Nick Abboud}
\thanks{These authors contributed equally to this work.}
\affiliation{Illinois Center for Advanced Studies of the Universe and Department of Physics, The Grainger College of Engineering, University of Illinois Urbana-Champaign, Urbana, Illinois 61801, USA}

\author{Yinchuan Lv} 
\affiliation{Department of Physics, The Grainger College of Engineering, University of Illinois Urbana-Champaign, Urbana, Illinois 61801, USA}
\affiliation{Materials Research Laboratory, The Grainger College of Engineering, University of Illinois Urbana-Champaign, Urbana, Illinois 61801, USA}

\author{Penghao Zhu}
\affiliation{Department of Physics, The Grainger College of Engineering, University of Illinois Urbana-Champaign, Urbana, Illinois 61801, USA}
\affiliation{Anthony J. Leggett Institute for Condensed Matter Theory, The Grainger College of Engineering, University of Illinois Urbana-Champaign, Urbana, Illinois 61801, USA}
\affiliation{Department of Physics, The Ohio State University, Columbus, OH 43210, USA}

\author{Azel Murzabekova} 
\affiliation{Department of Physics, The Grainger College of Engineering, University of Illinois Urbana-Champaign, Urbana, Illinois 61801, USA}
\affiliation{Materials Research Laboratory, The Grainger College of Engineering, University of Illinois Urbana-Champaign, Urbana, Illinois 61801, USA}

\author{Changjun Lee} 
\affiliation{Department of Materials Science and Engineering, The Grainger College of Engineering, University of Illinois Urbana-Champaign, Urbana, Illinois 61801, USA}
\affiliation{Materials Research Laboratory, The Grainger College of Engineering, University of Illinois Urbana-Champaign, Urbana, Illinois 61801, USA}

\author{Emma A. Pappas} 
\affiliation{Department of Physics, The Grainger College of Engineering, University of Illinois Urbana-Champaign, Urbana, Illinois 61801, USA}
\affiliation{Materials Research Laboratory, The Grainger College of Engineering, University of Illinois Urbana-Champaign, Urbana, Illinois 61801, USA}

\author{Dominic Petruzzi} 
\affiliation{Department of Physics, The Grainger College of Engineering, University of Illinois Urbana-Champaign, Urbana, Illinois 61801, USA}
\affiliation{Materials Research Laboratory, The Grainger College of Engineering, University of Illinois Urbana-Champaign, Urbana, Illinois 61801, USA}

\author{Jason Y. Yan} 
\affiliation{Department of Physics, The Grainger College of Engineering, University of Illinois Urbana-Champaign, Urbana, Illinois 61801, USA}
\affiliation{Materials Research Laboratory, The Grainger College of Engineering, University of Illinois Urbana-Champaign, Urbana, Illinois 61801, USA}

\author{Dipanjan Chauduri} 
\affiliation{Department of Physics, The Grainger College of Engineering, University of Illinois Urbana-Champaign, Urbana, Illinois 61801, USA}
\affiliation{Materials Research Laboratory, The Grainger College of Engineering, University of Illinois Urbana-Champaign, Urbana, Illinois 61801, USA}

\author{Peter Abbamonte} 
\affiliation{Department of Physics, The Grainger College of Engineering, University of Illinois Urbana-Champaign, Urbana, Illinois 61801, USA}
\affiliation{Materials Research Laboratory, The Grainger College of Engineering, University of Illinois Urbana-Champaign, Urbana, Illinois 61801, USA}

\author{Daniel P. Shoemaker} 
\affiliation{Department of Materials Science and Engineering, The Grainger College of Engineering, University of Illinois Urbana-Champaign, Urbana, Illinois 61801, USA}
\affiliation{Materials Research Laboratory, The Grainger College of Engineering, University of Illinois Urbana-Champaign, Urbana, Illinois 61801, USA}

\author{Rafael M. Fernandes} 
\affiliation{Department of Physics, The Grainger College of Engineering, University of Illinois Urbana-Champaign, Urbana, Illinois 61801, USA}
\affiliation{Anthony J. Leggett Institute for Condensed Matter Theory, The Grainger College of Engineering, University of Illinois Urbana-Champaign, Urbana, Illinois 61801, USA}

\author{Jorge Noronha} 
\affiliation{Illinois Center for Advanced Studies of the Universe and Department of Physics, The Grainger College of Engineering, University of Illinois Urbana-Champaign, Urbana, Illinois 61801, USA}

\author{Fahad Mahmood}
\affiliation{Department of Physics, The Grainger College of Engineering, University of Illinois Urbana-Champaign, Urbana, Illinois 61801, USA}
\affiliation{Materials Research Laboratory, The Grainger College of Engineering, University of Illinois Urbana-Champaign, Urbana, Illinois 61801, USA}

\maketitle

\noindent \textbf{In a system of charged chiral fermions driven out of equilibrium, an electric current parallel to the magnetic field can generate a dynamic instability by which electromagnetic waves become amplified.
Whether a similar instability can occur in chiral solid-state systems remains an open question.
Using time-domain terahertz (THz) emission spectroscopy, we detect signatures of what we dub a ``dynamic magneto-chiral instability" in elemental tellurium, a structurally chiral crystal. 
Upon transient photoexcitation in a moderate external magnetic field, tellurium emits THz radiation consisting of coherent modes that amplify over time. 
An explanation for this amplification is proposed using a theoretical model based on a dynamic instability of electromagnetic waves interacting with infrared-active oscillators of impurity acceptor states in tellurium to form an amplifying polariton. 
Our work not only uncovers the presence of a magneto-chiral instability but also highlights its promise for THz-wave amplification in chiral materials.}
\bigskip

Magnetic fields can induce anomalous, parity-breaking phenomena in chiral plasmas, i.e., systems with an inequivalent number of left- and right-handed chiral particles. 
A well-known example of such phenomena is the chiral magnetic effect (CME), where the existence of a chiral anomaly generates an electric current $\boldsymbol{j}$ parallel to an applied magnetic field $\boldsymbol{B}$, $\boldsymbol{j} =\sigma_{\text{M}}\boldsymbol{B}$ ~\cite{fukushima_chiral_magnetic_effect}. 
Here, $\sigma_{\text{M}}$ is a pseudoscalar that is odd under inversion and even under time reversal, reflecting the symmetries of the current $\boldsymbol{j}$ and the magnetic field $\boldsymbol{B}$. 
The CME is predicted in quark-gluon plasmas and can lead to a dynamic instability that amplifies magnetic fields as a function of time~\cite{PhysRevLett.111.052002, PhysRevD.90.125031,shovkovy2023anomalous}. Such a chiral plasma instability has garnered interest across various research fields such as early universe physics~\cite{early_universe}, supernovas~\cite{magnetar}, turbulence-driven dynamos in the post-merger phase of binary neutron star coalescence \cite{Most:2023sme}, and quark-gluon plasma studies~\cite{quark_gluon_plasma}.  
In solid state systems, the CME is evidenced in transport measurements on Weyl semi-metals ~\cite{Na3Bi,zhang2020magnetotransport}, and the associated dynamic instability is predicted to manifest as an anomalous reflectance of light from the surface of a time-reversal-breaking Weyl semi-metal prepared in a non-equilibrium steady state \cite{chiral_light_amplifier}.

\begin{figure*}
	\centering
	\includegraphics{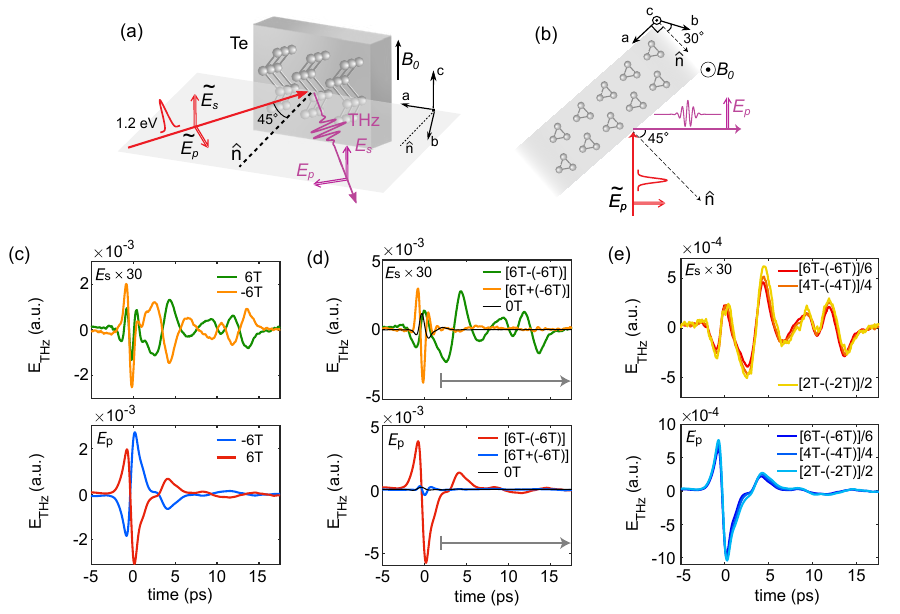}
	\caption{
		\textbf{THz emission from tellurium in an external magnetic field.} 
		\textbf{a}(\textbf{b}) The side (top) view of the sample geometry. The helical axis $\textbf{\textit{c}}$ is oriented parallel to the sample surface, which corresponds to the natural cleavage plane ($10\bar{1}0$) of tellurium. 
		The near-infrared (NIR) pump is incident onto the sample surface at 45$^{\circ}$. 
		The THz emission is collected at 90$^{\circ}$ from the incident NIR pump and is detected as either $\boldsymbol{E}_s$ ($\parallel \boldsymbol{B}_0$) or $\boldsymbol{E}_p$ ($\perp \boldsymbol{B}_0$).
		The sample surface normal is denoted with a dashed arrow $\hat{\textbf{n}}$, the rotation around which allows access to $\boldsymbol{B}_0\parallel\textbf{c}$ (shown in \textbf{a}, \textbf{b}) and $\boldsymbol{B}_0\perp\textbf{c}$ sample orientations.
		\textbf{c} $\boldsymbol{E}_s$ and $\boldsymbol{E}_p$ for $\boldsymbol{B}_0 = \pm 6$~T  at a temperature of 17~K. 
		\textbf{d} The $\boldsymbol{B}_0$-symmetrized and $\boldsymbol{B}_0$-antisymmetrized THz emission, along with the THz emission at 0~T. 
		The gray arrows mark the time window in which the 0~T and $\boldsymbol{B}_0$-symmetrized emission data are almost negligible but the $\boldsymbol{B}_0$-antisymmetrized emission data still show coherent oscillations.
		\textbf{e} $\boldsymbol{B}_0$-antisymmetrized data re-scaled by dividing the magnetic field amplitude $|\boldsymbol{B}_0|$.}  
	\label{fig:general_procedure}
\end{figure*}

Going beyond the CME in chiral plasmas, we propose that magneto-chiral currents $\boldsymbol{j}=\sigma_{\text{M}}\boldsymbol{B}$ in structurally chiral crystals driven out of equilibrium may also give rise to a dynamic instability.
$\sigma_{\text{M}}$, which transforms like a pseudoscalar (or an electric toroidal monopole), is symmetry-allowed in chiral crystals~\cite{kusunose2024emergence}. 
Although in equilibrium $\sigma_{\text{M}}$ must vanish according to band theory~\cite{axion_response_Burkov2013}, a finite-frequency response is still possible~\cite{GME_shudan2016}.
Furthermore, if an excitation generates an out-of-equilibrium unequal occupation of left-handed and right-handed electronic states in a transient state that is sufficiently long-lived, a magneto-chiral current analogous to the CME can arise.
Note that the imbalance between left-handed and right-handed electronic states can exist in chiral crystals even without the presence of emergent Weyl fermions at low energies.
As we show in this work, in the presence of electromagnetic waves, the out-of-equilibrium crystal with nonzero $\sigma_{\text{M}}$ can display a dynamic magneto-chiral instability manifested as an amplified response.

\begin{figure}
	\centering
	\includegraphics{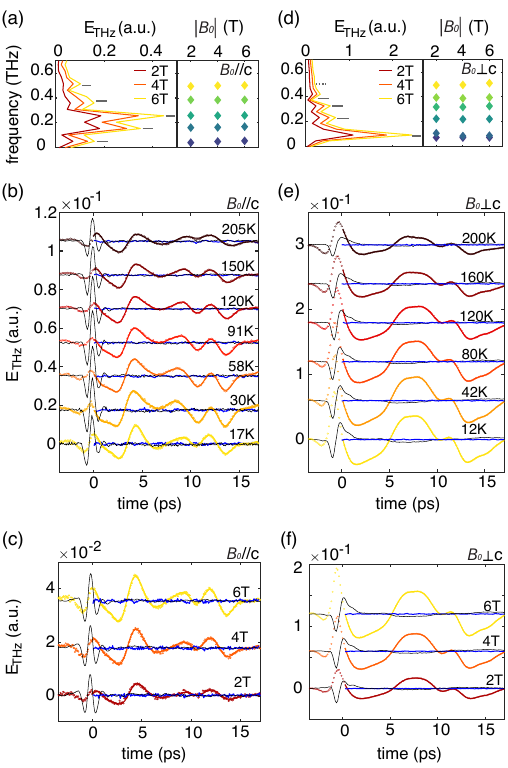}
	\caption{
		\textbf{Temperature and field-dependence of the $\boldsymbol{E}_s$ polarized THz emission $S_{\mathrm{odd}}$.}
		\textbf{a}-\textbf{c} Data for the $\boldsymbol{B}_0 \parallel \textbf{c}$ sample orientation:  
		\textbf{a} Fourier transform of the $\boldsymbol{E}_s$ emission $S_{\mathrm{odd}}$, with the $|\boldsymbol{B}_0|$ dependence of the extracted mode frequencies summarized in the right panel. 
		Modes are color-coded based on their frequency values.
		\textbf{b} Fit to the temperature-dependent $\boldsymbol{E}_s$ emission $S_{\mathrm{odd}}$ at 6~T. Colored circles represent the data, colored lines are the time-domain fits, and blue traces show the residuals. Black lines represent the $\boldsymbol{B}_0$-symmetrized emissions, providing a reference for $t=0$.  
		\textbf{c} Fit to the $|\boldsymbol{B}_0|$-dependent $\boldsymbol{E}_s$ emission $S_{\mathrm{odd}}$ at 17~K.
		\textbf{d}-\textbf{f} Data for the $\boldsymbol{B}_0 \perp \textbf{c}$ sample orientation:  
		\textbf{d} Fourier transform of the $\boldsymbol{E}_s$ emission $S_{\mathrm{odd}}$, and the $|\boldsymbol{B}_0|$ dependence of the extracted mode frequencies.
		\textbf{e} Fit to the temperature-dependent $\boldsymbol{E}_s$ emission data at 6~T.  
		\textbf{f} Fit to the $|\boldsymbol{B}_0|$-dependent $\boldsymbol{E}_s$ emission $S_{\mathrm{odd}}$ at 12~K.
		In \textbf{a} and \textbf{d} left panels, gray bars denote the linear-prediction-extracted mode frequencies averaged over different fields, with the one dashed bar indicating uncertainty due to the mode being near the noise level. }  
	\label{fig:resonance}
\end{figure}

We use time-domain THz emission spectroscopy in a magnetic field to reveal signatures of the proposed magneto-chiral dynamic instability in elemental tellurium, a nonmagnetic structurally-chiral crystal. 
Tellurium features helical atomic chains along its chiral axis $\textbf{c}$ [see Fig.~\ref{fig:general_procedure}(a-b)], allowing nonreciprocal transport \cite{gate_tunable} and current-induced magnetization~\cite{current_incudced_mag}.
We photoexcite tellurium in an external magnetic field $\boldsymbol{B}_0$ using an ultrafast near-infrared (NIR) pulse with energy 1.2~eV (greater than the 0.32~eV band-gap of semiconducting tellurium~\cite{te_Bandgap}). 
The photoexcitation generates a transient current and coherent modes that radiate THz waves, the electric field of which we resolve in the time domain using standard electro-optic sampling (Methods). 
We find that the radiated THz polarized along $\boldsymbol{B}_0$ contains coherent modes that unexpectedly increase in amplitude over time, consistently across varying magnetic fields and temperatures. 
We conjecture that this amplification is a signature of a dynamic instability in photoexcited tellurium at frequencies well below the band gap and approaching the hydrodynamic limit for charge carriers. 
To justify this conjecture, we develop a theoretical model in which the magneto-chiral instability is manifested in the polariton consisting of THz light coupled with an IR-active oscillator associated with a tellurium impurity level to form a polariton, leading to an amplifying, radiative instability that explains the observed THz amplification.

\begin{figure*}
	\centering
	\includegraphics{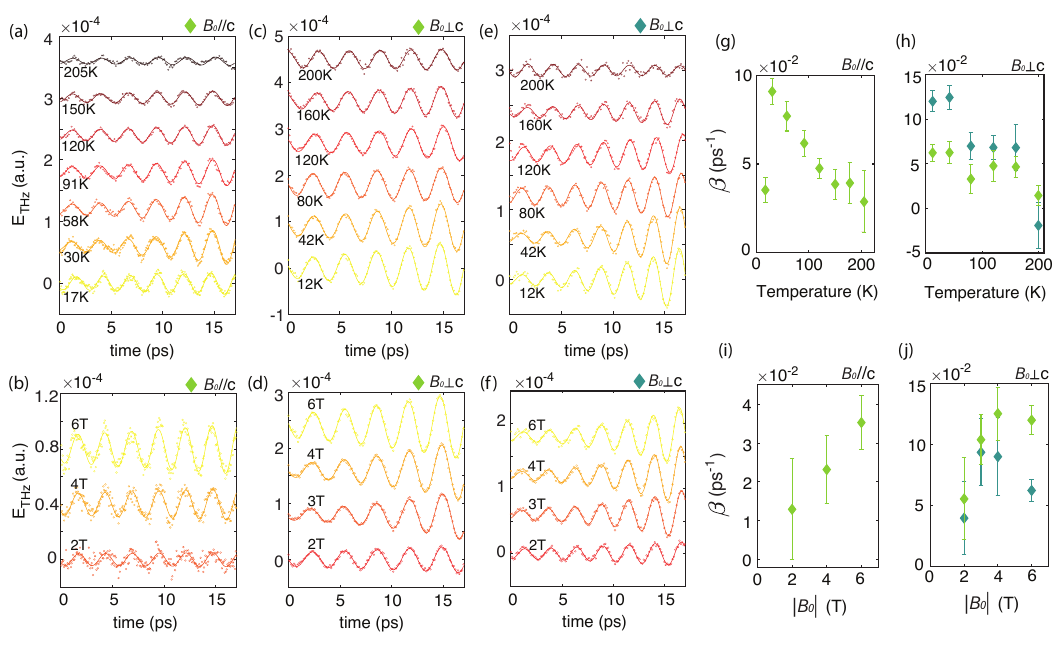}
	\caption{
		\textbf{The amplified modes in the $\boldsymbol{E}_s$ THz emission $S_{\mathrm{odd}}$.} 
		\textbf{a}-\textbf{b} $\boldsymbol{E}_s$-polarized $S_{\mathrm{odd}}$, with $\boldsymbol{B}_0 \parallel \textbf{c}$.
		\textbf{a} Temperature dependence of the 0.37~THz mode at 6~T. 
		\textbf{b} Same as \textbf{a}, but showing the $|\boldsymbol{B}_0|$ dependence at 17~K.
		\textbf{c}-\textbf{f} $\boldsymbol{E}_s$-polarized $S_{\mathrm{odd}}$ with $\boldsymbol{B}_0 \perp \textbf{c}$.
		\textbf{c} Temperature dependence of the 0.32~THz mode at 6~T.
		\textbf{d} Same as \textbf{c}, but showing $|\boldsymbol{B}_0|$ dependence at 12~K.
		\textbf{e} Temperature dependence of the 0.39~THz mode for $S_{\mathrm{odd}}$ taken at 6~T.
		\textbf{f} Same as \textbf{e}, but showing $|\boldsymbol{B}_0|$ dependence at 12~K.
		In \textbf{a}-\textbf{f}, colored circles represent the data after subtracting time-domain fit components of all other modes, and colored lines show the fits for the modes using parameters obtained from the linear prediction model.
		\textbf{g}, \textbf{h} Temperature dependence of the amplification rate $\beta$ for $\boldsymbol{B}_0 \parallel \textbf{c}$ and $\boldsymbol{B}_0 \perp \textbf{c}$, respectively.
		Color-coding of the modes for a given sample orientation is consistent with symbols on the upper right of \textbf{a}-\textbf{f} and with the color-coding in Fig.~\ref{fig:resonance} \textbf{a} and \textbf{d}. 
		For $\boldsymbol{B}_0 \perp \textbf{c}$, 0.32~THz (0.39~THz) is denoted with dark green (light green) symbols. 
		\textbf{i},\textbf{j} $|\boldsymbol{B}_0|$ dependence of the amplification rate $\beta$ for $\boldsymbol{B}_0 \parallel \textbf{c}$ and $\boldsymbol{B}_0 \perp \textbf{c}$, respectively. 
		The error bars in panels \textbf{g}-\textbf{j} are obtained from an additional least square fit using the output of the linear prediction model as initial parameters. 
	}
	\label{fig:amplified_modes}
\end{figure*}

Figure~\ref{fig:general_procedure}a,b show the side view and top view of the experimental geometry, respectively.
The helical axis $\textbf{c}$ of tellurium lies in the sample plane, and the NIR pulse is incident at 45$^{\circ}$ to the sample normal $\hat{\textbf{n}}$. 
The emitted THz is collected at 90$^{\circ}$ away from the incident NIR. Fig.~\ref{fig:general_procedure}c shows the measured time profiles of the THz electric field (${E}_\text{THz}(t)$) emitted from Te upon photoexcitation at a temperature of $T$ = 7~K and an external magnetic field $\boldsymbol{B}_0\parallel\textbf{c}$ of $\pm$6~T. 
Here the incident NIR pump is either $s$-polarized (electric field component $\tilde{\boldsymbol{E}}_s$ along $\boldsymbol{B}_0$) or $p$-polarized (electric field component $\tilde{\boldsymbol{E}}_p$ perpendicular to $\boldsymbol{B}_0$). 
Note that experiments were performed for two sample orientations: $\boldsymbol{B}_0\parallel\textbf{c}$ (shown in Fig.~\ref{fig:general_procedure}) and $\boldsymbol{B}_0\perp\textbf{c}$. 
To study the response that is odd or even with respect to the external magnetic field, we show the $\boldsymbol{B}_0$-antisymmetrized THz emission [(+6~T) data subtracted by (-6~T) data] and the $\boldsymbol{B}_0$-symmetrized [sum of (+6~T) and (-6~T) data], together with the 0~T THz emission in Fig.~\ref{fig:general_procedure}d. 
The sharp signal at $t = 0$~ps lasting for about $1$~ps observed in $\boldsymbol{B}_0$-symmetrized and the 0~T THz emission scan can be attributed to a photo-Dember effect from electron-hole mobility differences near the sample surface~\cite{DekorsyPhotodamberTellurium}. 
Strikingly, the $\boldsymbol{B}_0$-antisymmetrized data shows coherent oscillations that persist for a much longer time (shown for up to 17.5~ps in Fig.~\ref{fig:general_procedure}d) than typical THz emission due to just a photo-Dember effect. 
In contrast note that the 0~T and $\boldsymbol{B}_0$-symmetrized data show no THz emission after about 2~ps (time-window indicated by the gray arrow in Fig.~\ref{fig:general_procedure}d). 
The long-lasting signal in the $\boldsymbol{B}_0$-antisymmetrized data is also clearly evident in the raw THz emission from $s$-polarized pump (top panel, Fig.~\ref{fig:general_procedure}c). 
We further note that the $\boldsymbol{B}_0$-antisymmetrized emission at all times scales nearly linearly with the applied external magnetic field $\boldsymbol{B}_0$: scaling the emitted $\boldsymbol{B}_0$-antisymmetrized time profiles with $\boldsymbol{B}_0$ results in all curves nearly overlapping, as shown in Fig.~\ref{fig:general_procedure}e. 

Possible reasons for the increase in THz emission with increasing $\boldsymbol{B}_0$ include a radiating Hall current or rotation of a surface dipole radiation by an external magnetic field
~\cite{Johnston2002,shan2001origin}. 
These effects will enhance the THz emission \emph{only} in the $\boldsymbol{E}_p (\perp \boldsymbol{B}_0$) direction~\cite{Johnston2002,shan2001origin}, as the Lorentz force is always perpendicular to $\boldsymbol{B}_0$. 
For the remainder of the work, we thus focus on the $\boldsymbol{E}_s (\parallel \boldsymbol{B}_0$) anti-symmetrized emission (referred to as $S_{\mathrm{odd}}$ in the remainder of the text), which is not affected by the transport Hall current.

Figure~\ref{fig:resonance}a shows the Fourier transform of $S_{\mathrm{odd}}$ for $\boldsymbol{B}_0 \parallel \textbf{c}$. 
This frequency domain data consists of five main peaks, whose frequencies do not change significantly with magnetic field. 
Based on this observation, we apply a non-iterative time-domain linear prediction algorithm~\cite{barkhuijsen1985retrieval} to the time-domain THz data $S_{\mathrm{odd}}$. 
Linear prediction decomposes the data into a sum of harmonic oscillators with exponential factors $\sum_i A_i e^{\beta_i t}\cos{(\omega_i t+\phi_i)}$.
This procedure requires no initial fit parameters and uniquely fits noise-free sinusoidal data with exponential envelopes (SI Sec.A), whereas for noisy data, it reliably identifies sinusoidal components above the noise level, as shown in NMR~\cite{led1991application} and pump-probe spectroscopy~\cite{Yijing_3}. 
We find that constraining the fitting algorithm for $S_{\mathrm{odd}}$ to allow only decaying exponentials ($\beta<0$) does not fit the data well - there are significant residuals with amplitudes increasing with time (SI Fig.~S1).
In contrast, allowing for growing exponentials ($\beta>0$) accurately describes both the magnetic field and temperature dependent $S_{\mathrm{odd}}$ data as shown in Fig.~\ref{fig:resonance}b-c. 
The time-domain linear prediction fits (colored solid lines) closely match the data (empty circles), with negligible residuals (blue solid lines). 
We performed the same analysis for the $S_{\mathrm{odd}}$ signal measured for $\boldsymbol{B}_0 \perp \textbf{c}$, and find that allowing $\beta>0$ is necessary for obtaining fits that accurately describe the data, as shown in Fig.~\ref{fig:resonance}e-f. 
From the fitting, we extract five distinct sinsusoidal modes for both sample orientations. 
The extracted frequencies of these modes are shown in the right panels of Fig.~\ref{fig:resonance}a and Fig.~\ref{fig:resonance}d. 
As expected from the scaling behavior of the time-domain data (Fig.~\ref{fig:general_procedure}e), there is negligible dependence of the mode frequencies on the external magnetic field. 
We further note two caveats to the extracted modes shown in Fig.~\ref{fig:resonance}d: the small splitting of the $<0.1$~THz mode is attributed to mode chirp (SI Sec.\ A); and the amplitude of the highest-frequency mode is close to the noise level.
These caveats do not affect the main conclusions of this work.

Having developed a reliable method to extract the dynamics of the modes contributing to $S_{\mathrm{odd}}$, we next determine their possible origin. 
Since these modes radiate in the far-field, they must have a net infrared-dipole moment and, given that the crystal structure of tellurium breaks inversion symmetry, suggesting the modes may originate from optical phonons. 
However, the frequencies of the observed modes are much smaller than any of tellurium’s optical phonons~\cite{DekorsyPhotodamberTellurium,torrie1970raman}. 
Moreover, the absence of coherent oscillations in a 1030~nm pump~-~515~nm transient reflectivity probe measurement under $\boldsymbol{B}_0\parallel\textbf{c}$ (SI Sec.\ B) indicates that they are not Raman-active. 
Instead, the energies of the observed modes are close to the energies of acceptor states in tellurium due to impurities (e.g. selenium or group V elements)~\cite{Couder1973,BetbederMatibet1969,Tani1979,natori1973acceptor} as measured previously in infrared magneto-transmission experiments. Those results also showed negligible $\boldsymbol{B}_0$-field dependence of the energies~\cite{Hardy1971,Couder1973,shinno1973conduction}, consistent with our observations (SI Sec.\ C).
The mode frequencies differ between the two sample orientations (gray bars in Fig.~\ref{fig:resonance}b-c), likely due to the anisotropic nature of tellurium.

\begin{figure}
	\centering\includegraphics{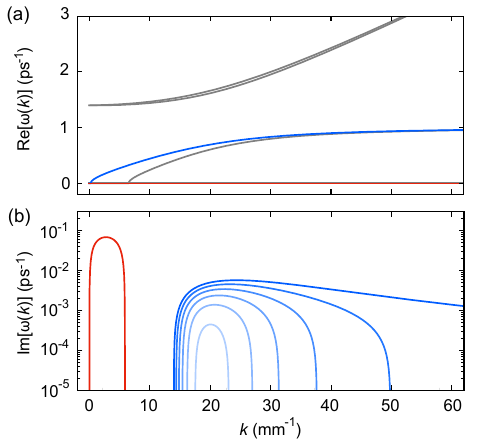}
	\caption{
		\textbf{Calculated dispersions of the polariton illustrating the magneto-chiral dynamic instability.} 
		\textbf{a} The $\operatorname{Re}[\omega(k)]$ of the plane-wave solutions. 
		Gray: Stable modes ($\operatorname{Im}[\omega(k)]<0$) with an avoided crossing, similar to the conventional polariton resulting from an IR-oscillator coupling with a photon. 
		Red: A non-propagating $(\operatorname{Re}[\omega(k)]=0)$ instability of the electromagnetic wave. 
		Blue: The amplyifing polariton with $\operatorname{Re}[\omega(k)]$ pinned to $\omega_0$, assuming $\gamma=0$. The  $\operatorname{Re}[\omega(k)]$ dispersions are mostly independent of $\gamma$.
		\textbf{b} The $\operatorname{Im}[\omega(k)]$ of the solutions. 
		Red: $\operatorname{Im}[\omega(k)]$  of the non-radiating instability. 
		Blue: $\operatorname{Im}[\omega(k)]$ of the amplifying polariton assuming different $\gamma$. From dark blue to light blue, $\gamma$ increases from 0 to $2.5\times10^{-2}\ \text{ps}^{-1}$ in increments of $5\times10^{-3}\ \text{ps}^{-1}$. 
	}  
	\label{fig:dispersions}
\end{figure}

We next focus on the modes in $S_{\mathrm{odd}}$ that are described by exponentially growing sinusoids ($\beta>0$) and investigate their temperature and field dependence. 
We isolate the $\beta>0$ modes by subtracting the time-domain fits of all the other modes.
In the $\boldsymbol{B}_0\parallel\textbf{c}$ measurement geometry, 
Fig.~\ref{fig:amplified_modes}a-b show the results for the 0.37~THz $\beta>0$ mode. 
In the $\boldsymbol{B}_0\perp\textbf{c}$ geometry, Fig.~\ref{fig:amplified_modes}c-d show the results for the 0.32~THz $\beta>0$ mode and Fig.~\ref{fig:amplified_modes}e-f for the 0.39~THz $\beta>0$ mode.
The isolated mode trajectories (empty circles) overlap relatively well with the subtracted trajectories (solid lines).  
The dependence of the extracted $\beta$ on the applied field and temperature is summarized in Fig.~\ref{fig:amplified_modes}g-j, which shows that $\beta>0$ consistently for these modes except at high temperature. 

We now provide an interpretation of the amplifying components in the $S_\text{odd}$ emissions by developing a model based on magneto-chiral currents.
In general, the free-electron current density $\bm j$ can be written as follows, assuming linear response on the Fourier domain of time and space:
\begin{align} \label{j}
	\bm j(\omega, \bm k) = \sigma_{\text{M}}(\omega) \bm B(\omega, \bm k)
	+ \sigma_{\text{E}}(\omega) \bm E(\omega, \bm k)
\end{align}
where $\boldsymbol{E}$ and $\boldsymbol{B}$ refer to the total electric and magnetic fields, respectively, as a function of the angular frequency $\omega$ and wave-vector $\bm k$, and $\sigma_{\text{E}}$ is the electric conductivity. 
The pseudoscalar $\sigma_{\text{M}}$ describes the frequency-dependent magneto-chiral conductivity. Importantly, $\sigma_{\text{M}}(\omega\rightarrow0)$ vanishes in a near-equilibrium situation but can be nonzero for a non-equilibrium carrier distribution between electronic states with opposite chirality~\cite{GME_shudan2016}. 
Thus, in the $\bm k \to 0$ limit, we assume that $\sigma_{\text{M}}(\omega)$ for a crystal driven out of equilibrium includes both adiabatic and frequency-dependent responses,
\begin{align} \label{eq:sigmaM}
	\sigma_{\text{M}}(\omega) &= \frac{e^2}{4\pi^2 \hbar^2} \left[ \mu_{\text{DC}} + \mu_{\text{AC}} \frac{\omega \tau}{\omega \tau + i} \right],
\end{align}
where $\mu_{\text{AC}}$ and $\mu_{\text{DC}}$ are real-valued parameters with dimensions of energy.
This form satisfies causality in linear response. 
The frequency dependence of the $\mu_{\text{AC}}$ term can be derived from the Boltzmann equation in the single relaxation-time approximation~\cite{GME_shudan2016}, for which $\tau$ denotes the elastic scattering lifetime of free carriers.
We emphasize that small perturbations near equilibrium do not give rise to a finite $\mu_{\text{DC}}$, which instead represents a far-from-equilibrium effect. 
By assuming constant $\mu_{\text{DC}}$ and $\mu_{\text{AC}}$, we implicitly consider the system to be in a long-lived non-equilibrium transient state.
This assumption is justified since, in our experiment, photoexcitation of tellurium causes a long-lived non-equilibrium carrier distribution in tellurium’s conduction and valence bands, persisting over the time window shown in Figs.~1-3. 
Note that beyond this time-window ($>30$~ps), the THz emission data indicates that $\sigma_{\text{M}}$ decays (i.e., $\mu_{\text{AC}}$ and $\mu_{\text{DC}}$ decay) between 20-30~ps (SI Sec.\ D), consistent with the carrier recombination time~\cite{jnawali2020ultrafast}.

Given that the energies of the observed modes correspond to energies of impurity acceptor states, we model $\sigma_{E}$ as a sum of the DC conductivity $\sigma_{E0}$ and a contribution from the bound current, 
\begin{align} \label{eq:sigmaE}
	\sigma_{\text{E}}(\omega) = \sigma_{E0} - \frac{i \omega C}{\omega_0^2 - \omega^2 - i \gamma \omega}.
\end{align}
The second term is the response of an infrared (IR)-active oscillator with intrinsic angular frequency $\omega_0$, damping rate $\gamma$, and  $C = n e^2/m$ where $n$ is the impurity concentration and $m$ the oscillator mass~\cite{dresselElectrodynamicsSolidsOptical2002}.

Inserting Eqs.~(1-3) into Maxwell's equations and using a plane-wave ansatz for electromagnetic fluctuations, $\bm E, \bm B \propto e^{i(\bm k \cdot \bm r - \omega t)}$, we calculate the dispersion relations $\omega(\bm k)$ numerically.
We use physically reasonable parameters: $\sigma_{E0} = 50\ \Omega^{-1}\ \text{m}^{-1}$ \cite{bottom}, $\omega_0 = 1\ \text{ps}^{-1}$ (on the order of the observed mode frequencies), $n = 10^{16}\ \text{cm}^{-3}$, $\mu_{\text{DC}} = 0.5\ \text{eV}$, $ \mu_{\text{AC}} = -0.5\ \text{eV}$, $\tau = 1\ \text{ps}$, and the relative permittivity $\epsilon_r = 30$. 
Note that $ \mu_{\text{AC}}$ and $\mu_{\text{DC}}$ approximate the energy difference between the pump laser photon (1.2~eV) and tellurium’s band gap (0.32~eV). 
The results are shown in Fig.~\ref{fig:dispersions}a for the real part $\operatorname{Re}[\omega(k)]$ and in Fig.~\ref{fig:dispersions}b for the imaginary part $\operatorname{Im}[\omega(k)]$, with both depending only on $k=|\bm k|$ due to the model isotropy.
Two unstable solutions ($\operatorname{Im}[\omega(k)]>0$, red and blue) are present, implying that these electromagnetic modes grow as $e^{\operatorname{Im}[\omega(\bm k)] t}$. 
We associate these modes to the magneto-chiral dynamical instability. 
The stable modes ($\operatorname{Im}[\omega(k)]<0$, gray lines in Fig.\ \ref{fig:dispersions}a) are the conventional damped polariton arising due to the coupling between the IR-active mode and the electromagnetic wave~\cite{huang1951interaction}. The slight degeneracy lift is due to helicity-breaking $\sigma_{\mathrm{M}}$.
One of the two instabilities shown in Fig.~\ref{fig:dispersions}  (red) is non-radiative with vanishing group velocity and Poynting vector. 
This instability resembles the chiral plasma instability discussed in Ref.~\cite{PhysRevLett.111.052002} and occurs due to $\mu_{\text{DC}}\neq0$. 
Adding a Hall current under an external magnetic field enables propagation of this red instability, reproducing the chiral magnetic instability noted in Refs.~ \cite{chiral_light_amplifier,PhysRevB.107.014302}. 
However, an unrealistically large value of $\mu_{\text{DC}}$ would be required to observe this red instability in the THz frequency range of the oscillators in $S_{\text{odd}}$ (SI Sec.\ E).

In contrast, the second magneto-chiral instability (blue in Fig.~\ref{fig:dispersions}) is radiative ($\operatorname{Re}[\omega(k)]\neq0$) for the physically reasonable parameters noted above.
Moreover, the angular frequency is pinned at $\operatorname{Re}[\omega(k)] \approx \omega_0 = 1\,\mathrm{ps}^{-1}$ for $k$ larger than $ 25$~mm$^{-1}$, which is the value that maximizes $\operatorname{Im}[\omega(k)]$. 
This frequency pinning arises from the polaritonic coupling, where the IR-active oscillator gains instability through its interaction with the electron fluid via electromagnetic waves.
Such polaritonic coupling via the magneto-chiral instability can be generalized to other IR-active oscillators (e.g. optical phonons).
We note that the generation of collective modes with large wave vectors up to $k\sim 100$~mm$^{-1}$ is allowed given the short laser penetration depth (300 nm~\cite{tutihasi1969optical}, considerably shorter than the emitted THz wavelength).   
While we do not track the time evolution of $\mu_{\text{DC}}$ and $\mu_{\text{AC}}$ in the model, it is reasonable to expect that the energy to power the radiative instability due to a finite $\mu_{\text{AC}}$ is supplied by the decay of the non-zero $\mu_{\text{DC}}$, which represents the system relaxing from a far-out-of-equilibrium state.

As shown in Fig.~\ref{fig:dispersions}b, without damping ($\gamma=0$), the radiative instability's $\operatorname{Im}[\omega(k)]$ of approximately 10$^{-2}$~ps$^{-1}$ aligns within an order of magnitude with the observed $\beta$. 
As $\gamma$ increases, the amplification rate $\operatorname{Im}[\omega(k)]$ of the instability decreases. $\gamma$ is sensitive to disorder, phonon scattering, and local potentials in cyclotron motions, and it typically increases with temperature and decreases with magnetic field under lower fields~\cite{hopkins1989temperature}.
Thus, we expect the amplification rate to decrease (increase) with temperature (magnetic field at lower fields), which is indeed observed in the experimentally extracted $\beta$, see Fig.~\ref{fig:amplified_modes}g-j.
Note that if $\gamma$ exceeds $3\times10^{-2}\ \text{ps}^{-1}$, then the radiative instability vanishes ($\operatorname{Im}[\omega(k)]$ is no longer $> 0$), which provides a viable explanation for the experimental observation that only a subset of the IR-active modes are amplified over time (those with smaller $\gamma$).

We next discuss the crystal symmetry requirements for generating a magneto-chiral current with photoexcitation. 
In a crystal, the coefficient $\sigma_{\text{M}}$ is replaced by a rank-2 tensor $\sigma_{\text{M},ij}$ that relates the current component $j_i$ with the magnetic field component $B_j$, as discussed in Ref~\cite{GME_shudan2016}. 
The non-zero elements of this tensor for a given crystal structure can be derived via group theory. 
Using the facts that Te has the chiral space group $P3_121$ with corresponding point group $D_3$, and that the out-of-plane (in-plane) components of both the current and the magnetic field transform as the same non-trivial irreducible representation $A_2$ ($E$) of $D_3$, it follows that $\sigma_{\text{M},ij}$ is a diagonal tensor with $\sigma_{\text{M},xx}=\sigma_{\text{M},yy}\neq\sigma_{\text{M},zz}$~\cite{PhysRevResearch.2.012073}.
This reflects the fact that a pseudoscalar transforms as the trivial irreducible representation, as expected for any of the 11 enantiomorphic point groups~\cite{anastassakis1980morphic}. 
However, symmetry alone is insufficient to ensure a finite current, since $\mu_{\text{DC}}$ is only non-zero in a far-from-equilibrium situation where electronic states with opposite chirality have different occupations.
In our experiment, we expect that after the initial photoexcitation, charge fluctuations that preserve the crystalline symmetries (i.e., that transform as the $A_1$ irreducible representation) will be the long lived ones~\cite{A1force_2013, A1force_2019}. 
Since the pseudoscalar also transforms as the $A_1$ irreducible representation, the photoexcitation can potentially generate a long-lived non-equilibrium distribution of carriers with a net chirality.

We note that the microscopic origin of $\mu_{\text{DC}}$ and $\mu_{\text{AC}}$ depends on the properties of the underlying band structure, as discussed in \cite{GME_shudan2016}. 
While chiral-imbalanced Weyl nodes in tellurium have been previously reported to lead to a CME~\cite{zhang2020magnetotransport}, which can be described with a $\mu_{\text{DC}}$ term, the Weyl nodes are located a few hundreds of meV away from the middle of the band gap \cite{chang2018topological,Ma_Te_Weyl_band_Natcomm2022,zhang2020magnetotransport,chen2022topological,Hirayama2015}.    
Future studies of the non-equilibrium carrier distributions (e.g., using time-resolved ARPES experiments combined with DFT calculations) can provide further insights into the microscopic origin of the magneto-chiral dynamic instability in tellurium.
Importantly, we emphasize that our analysis relies only on the chiral symmetry of the crystal structure of tellurium, and not on the precise microscopic mechanisms at play. 
Our findings highlight the potential to manipulate THz-range electromagnetic waves in far-out-of-equilibrium chiral materials, including wave amplification.

\vspace{2mm}
\noindent\textbf{Acknowledgments:}
We thank Enrico Speranza, Rong Zhang, Yingkai Liu, Taylor Hughes, and Thomas P. Devereux for useful discussions.
This work was supported by the Quantum Sensing and Quantum Materials, an Energy Frontier Research Center funded by the U.S. Department of Energy (DOE), Office of Science, Basic Energy Sciences (BES), under Award No.DE-SC0021238 (project conception, data acquisition and analysis, and manuscript preparation).
F.M., Y.L., A.M., and C.L. acknowledge support from NSF Career Award No.\ DMR-2144256 for the development of the experimental setup.
F.M. acknowledges support from the EPiQS program of the Gordon and Betty Moore Foundation, Grant GBMF11069.
Y.H. acknowledges support from the IQUIST (Illinois Quantum Information Science and Technology Center) Postdoctoral Fellowship.
P.Z was supported by the Center for Emergent Materials, an NSF MRSEC, under award number DMR-2011876. 
R.M.F. (theoretical model) was supported by the Air Force Office of Scientific Research under Award No. FA9550-21-1-0423.

\vspace{2mm}
\noindent\textbf{Author contributions:} Y.H. performed sample characterization, the THz emission experiments, and the corresponding data analysis. 
Y.L., A.M., Y.H., C.L., and F.M. designed and built the THz emission setup compatible with a magnetic field.
D.P. and J.Y.Y. assisted with the THz emission measurements without the echo.
N.A., Y.H., P.Z., R.M.F., and J.N. developed the theoretical understanding and modeling of the magneto-chiral instability.  
E.A.P. performed preliminary synthesis reactions and assisted Y.H. with structural characterization, supervised by D.P.S.
D.C. and P.A. provided insights into the impurity levels in tellurium.
Y.H., N.A., P.Z., R.M.F., J.N., and F.M. wrote the manuscript with input from all the authors. 
F.M. conceived and supervised this project.

\vspace{2mm}
\noindent\textbf{Competing interests:} The authors declare no competing interests.

\vspace{2mm}
\noindent\textbf{Methods:} THz emission measurements were performed with our custom-built time-domain THz spectroscopy setup based on a Yb:KGW amplifier laser (PHAROS, Light Conversion). The fundamental laser pulse wavelength is 1030~nm with a pulse duration of $\sim160$ fs. The fundamental beam is split into a pump and a probe, with the pump incident onto the sample at a 45$^\circ$ angle-of-incidence while the probe is used for electro-optic sampling (EOS) of the emitted THz field. The sample is mounted at the center of a magneto-optic cryostat (OptiCool, Quantum Design) with TPX windows to allow THz transmission. The THz field radiated by the sample is collected and collimated at 90$^{\circ}$ from the pump beam and at a 45$^{\circ}$ angle relative to the sample normal, using a THz-transmissive TPX lens mounted inside the cryostat. The emitted THz is then focused onto a (110)-cut CdTe crystal. The EOS probe beam is made to spatially and temporally overlap with the emitted THz on the CdTe crystal. The THz field $\boldsymbol{E}~(t)$ is thus measured by scanning the time delay of the 1030~nm electro-optic sampling beam relative to the emitted THz field~\cite{planken2001measurement_EOsampling}. The detection bandwidth of the CdTe crystal decreases above 1~THz~\cite{ropagnol2020toticient}. The experiments were performed with two prototypical sample orientations, $\boldsymbol{B}_0\parallel\textbf{c}$ and $\boldsymbol{B}_0\perp\textbf{c}$, which can be accessed by rotating the sample. To selectively detect the THz emission component $\boldsymbol{E}_s$ parallel to the $\boldsymbol{B}$ field or the component $\boldsymbol{E}_p$ perpendicular to the $\boldsymbol{B}$ field, we rotated four optical components: the THz wire grid polarizer (WGP) with an extinction ratio $10^{-3}$, the half-wave plate, and polarizer for the 1030 nm probe, and the CdTe crystal for EO sampling. The WGP has an extinction ratio of better than 10$^{-3}$.

\vspace{2mm}
\noindent\textbf{Data availability:} The data in this manuscript is available at Illinois Data Bank~\cite{databank}.

\vspace{2mm}
\noindent\textbf{Corresponding authors:} Correspondence to Yijing Huang~(huangyj@illinois.edu) and/or Fahad Mahmood~(fahad@illinois.edu).

\bibliography{aTe}

\end{document}